# *Manikin-Recorded Cardiopulmonary Sounds Dataset Using Digital Stethoscope*

*Yasaman Torabi[1], Shahram Shirani[1,2], James P. Reilly[1]*

[1]Electrical and Computer Engineering Department, McMaster University, Hamilton, Ontario, Canada
[2] L.R. Wilson/Bell Canada Chair in Data Communications, Hamilton, Ontario, Canada

CORRESPONDING AUTHOR: Yasaman Torabi (e-mail: torabiy@mcmaster.ca).

**ABSTRACT** Heart and lung sounds are crucial for healthcare monitoring. Recent improvements in stethoscope technology have made it possible to capture patient sounds with enhanced precision. In this dataset, we used a digital stethoscope to capture both heart and lung sounds, including individual and mixed recordings. To our knowledge, this is the first dataset to offer both separate and mixed cardiorespiratory sounds. The recordings were collected from a clinical manikin, a patient simulator designed to replicate human physiological conditions, generating clean heart and lung sounds at different body locations. This dataset includes both normal sounds and various abnormalities (i.e., murmur, atrial fibrillation, tachycardia, atrioventricular block, third and fourth heart sound, wheezing, crackles, rhonchi, pleural rub, and gurgling sounds). The dataset includes audio recordings of chest examinations performed at different anatomical locations, as determined by specialist nurses. Each recording has been enhanced using frequency filters to highlight specific sound types. This dataset is useful for applications in artificial intelligence, such as automated cardiopulmonary disease detection, sound classification, unsupervised separation techniques, and deep learning algorithms related to audio signal processing.

## I. BACKGROUND

Cardiopulmonary diseases are significant contributors to global mortality rates. Chronic respiratory diseases, such as asthma, were responsible for over 147,000 deaths in 2022 alone [1]. Meanwhile, cardiovascular diseases remain the leading cause of death worldwide, accounting for approximately 18.6 million deaths annually [2]. Therefore, it is crucial to accurately analyze both heart and lung functions.

Auscultation of heart and lung sounds plays a vital role in diagnosing a variety of cardiopulmonary conditions [3]. Although these acoustic signals are weak, they hold essential medical information [4]. In 1816, René Laennec invented the first stethoscope to listen to body sounds, which has evolved over the years from a simple acoustic device to more sophisticated digital versions [5].

Technological advancements have led to the development of electronic stethoscopes that convert sound waves into electrical signals, allowing for sound processing and recording [6, 7]. Digital stethoscopes are the latest generation of these auscultation devices. They not only have the features of electronic stethoscopes, but also offer enhanced analysis capabilities, including integration with smartphones and cloud-based platforms for real-time analysis and sharing [8].

Artificial intelligence (AI) and machine learning algorithms have made significant improvements in real-time analysis and clinical decision-making [9, 10]. However, the effectiveness of these models heavily depends on the availability of high-quality datasets that contain diverse examples and enough dataset size. Such datasets are essential for training and validating AI models that perform sound classification, anomaly detection, and signal separation. Currently, there are few available datasets containing heart and lung sounds. The invention of patient simulators has become a critical point in clinical training and data collection, offering a risk-free and realistic environment for recording heart and lung sounds. For example, the manikin used in this work is widely used for its realistic articulation and ability to simulate a broad range of clinical scenarios [11].

In this work, we collected a total of 210 audio recordings (101 female and 109 male) from a clinical skills manikin in a controlled, noise-free environment. We recorded from 12 distinct chest locations, including standard heart and lung auscultation landmarks. Our dataset includes 50 heart sound recordings, 50 lung sound recordings, and 110 mixed recordings with ten heart sound types and six lung sound types. We controlled the manikin using an instructor tablet connected via Wi-Fi, allowing real-time adjustments. We recorded the sounds using a digital stethoscope, connected via Bluetooth to a cellphone application, which enabled real-time visualization, recording, and storage. Each 15-second recording was saved in *.wav* format and uploaded to a cloud-based platform for further analysis on a personal computer (PC), offering a valuable resource for studying cardiopulmonary sounds.

In 2019, Kun-Hsi Tsai *et al.* were the first to utilize a student auscultation manikin to record heart and lung sounds at 8 kHz sampling rate. However, the dataset was focused only on specific lung conditions (i.e., normal, wheezing, rhonchi, and stridor), and normal heart sound only. Moreover, they recorded heart and lung sounds separately and mixed them later [12]. Compared to Kun-Hsi Tsai *et al.*, we captured more diverse lung sound types, as well as heart abnormalities [12].

Later in 2021, Luay Fraiwan *et al.* recorded a dataset, focused solely on lung sounds [13]. Similarly, Jorge Oliveira *et al.* introduced the "*CirCor DigiScope Phonocardiogram*" dataset. They collected over 5,000 heart sound recordings but lacked lung sound data [14]. In contrast to Fraiwan *et al.* [13] and Oliveira *et al.* [14], who focused solely on lung sounds and heart sounds, respectively, our dataset includes heart, lung, and mixed recordings. In 2022, Julio Alejandro Valdez *et al.* gathered a cardiopulmonary dataset which provided separate heart and lung sound files recorded simultaneously but did not offer mixed recordings [15]. Unlike this work, we recorded mixed sound recordings, which are particularly valuable for developing unsupervised algorithms like blind source separation, allowing for analysis of the natural overlap between heart and lung sounds.

Compared to previous works, which used an electronic stethoscope, our work utilizes a digital stethoscope, providing improved recording quality, a higher sampling rate, and better resolution. Our dataset also offers better sound quality due to pre-recording filtering and covers a wider range of heart and lung abnormalities. If the patient simulator and stethoscope configuration are followed as described in this paper, the dataset is reusable in various scenarios and applications. A key advantage of using the manikin is its ability to provide clear, isolated recordings of heart-only, lung-only, as well as heart-lung mixture sounds recorded simultaneously in a risk-free and realistic environment.

This dataset is particularly valuable for developing automated machine-learning algorithms for detecting heart and lung diseases. It is ideal for biomedical engineering and AI researchers focused on disease detection, heart-lung sound classification, and signal processing related to audio data. This dataset can be useful for various signal processing algorithms, such as principal component analysis (PCA), independent component analysis (ICA), support vector machine (SVM), convolutional neural networks (CNNs), filtering, etc. The dataset is useful for both supervised and unsupervised learning methods. Distinct heart and lung sounds are useful for supervised methods, such as classification algorithms. For unsupervised methods, such as clustering or blind source separation, the mixture recordings offer a rich resource for exploring the separation of overlapping signals.

We utilized this dataset in signal processing and machine learning studies. In [16], we proposed a modified affine non-negative matrix factorization (NMF) method for blind separation of heart and lung sounds. This method used a parallel structure of multilayer units and exploited the periodic nature of heart and lung signals.

## II. COLLECTION METHODS AND DESIGN

### A. Setup and Environment

We performed auscultation in a controlled, quiet, and noise-free environment (Fig. 1A). We positioned the patient simulator in a sitting position to simulate realistic conditions for lung and heart sound recordings (Fig. 1B).

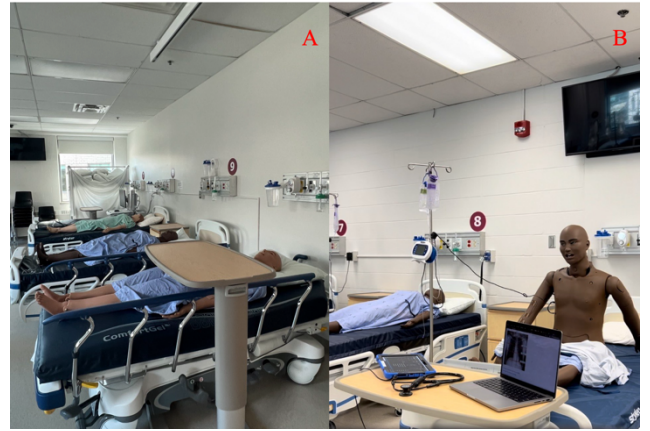

**FIGURE 1.** Setup and Environment: (A) Recording unit at Professional Practice Collaboratory (PPC), and (B) Manikin in a sitting position alongside the recording setup.

We recorded sounds from various chest locations depending on whether we were capturing heart or lung sounds (Fig. 2). For the mixed recordings, we selected the location from either heart or lung zones. We performed lung recordings from both the right and left sides of the chest, with each side divided into three zones: upper, middle, and lower. We focused on the anterior regions of the chest to ensure high-quality recordings, as the manikin's speakers are positioned at the front.

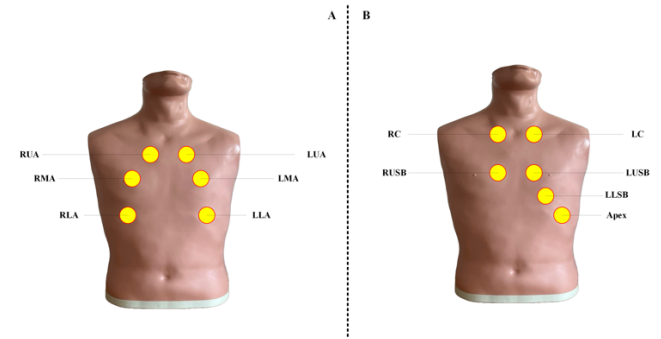

**FIGURE 2.** Chest zone landmarks for recording: (A) lung sounds, and (B) heart sounds.

We used standard lung auscultation landmarks [17]: Upper Anterior (UA), which is on ribs 2-4; Middle Anterior (MA), located at the anterior surface of ribs 4-6; and Lower Anterior (LA), placed over ribs 6-8, 45° down from nipple. For heart auscultation, we performed auscultation over the classic sites of auscultation [28] as follows:

- Apex (A): Mitral area
- Right Upper Sternal Border (RUSB): Aortic area
- Left Upper Sternal Border (LUSB): Pulmonary area
- Left Lower Sternal Border (LLSB): Tricuspid area
- Right Costal Margin (RC)
- Left Costal Margin (LC)

### B. Clinical Skills Manikin and Control System

We utilized the *CAE Juno™* nursing skills manikin, a mid-fidelity patient simulator designed to enhance clinical nursing skills. This manikin features interchangeable male and female chest skins, allowing gender switching to simulate male and female patients [19, 20].

We employed the *CAE Maestro* software, installed on a tablet, to control and monitor the manikin in real time. We connected the tablet to the manikin via a secure Wi-Fi connection, to remotely manage and customize a variety of patient sounds [21]. Fig. 3 shows the manikin and the tablet running the controlling software.

The Manikin offers a range of pre-recorded heart and lung sounds, which are synchronized with the cardiac cycle and ventilation of the left and right lungs, respectively [22]. Using the Maestro software, we adjusted the simulation parameters to suit the specific requirements of our study, simulating real-life auscultation scenarios at various chest locations.

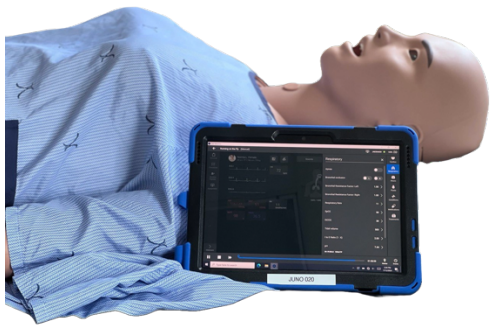

**FIGURE 3.** Clinical manikin and the instructor tablet.

First, we enabled only heart sounds and recorded them from the standard auscultation landmarks, generating various normal and abnormal sounds: normal, late diastolic murmur, mid systolic murmur, late systolic murmur, atrial fibrillation, s4 (fourth heart sound), early systolic murmur, s3 (third heart sound), tachycardia, and atrioventricular block. Next, we enabled only lung sounds and recorded the following types: normal, wheezing, crackles, rhonchi, pleural rub, and gurgling sounds. Finally, we enabled both heart and lung sounds and recorded them simultaneously. The different cases in the dataset follow a uniform distribution of recording location, sound type, and gender, ensuring the dataset closely simulates real-world scenarios and covers all possible outcomes.

### C. Digital Stethoscope and Recording

We used the *3M™ Littmann® CORE* Digital Stethoscope, 3M's most advanced model, to record lung and heart sounds (Fig. 4). Each recording lasted 15 seconds, which was sufficient to capture at least one complete respiratory or cardiac cycle. The stethoscope has built-in frequency filters, allowing us to selectively capture heart sounds, lung sounds, or both, thereby enhancing precision during the recording process. We applied three different filter modes based on the type of sound being captured: Bell mode for recording low-frequency heart sounds, Diaphragm mode for capturing high-frequency lung sounds, and Midrange mode for recording both heart and lung sounds simultaneously. The amplification feature provided up to 40x sound enhancement, and the active noise cancellation effectively reduced ambient noise [23, 24].

We connected the stethoscope to the *Eko* software via Bluetooth, which enabled us to store the recordings directly on a mobile device. This mobile application pairs with the digital stethoscope, to record, visualize, and analyze cardiorespiratory sounds. It allows real-time monitoring, playback, and cloud-based storage of the recorded sounds. These recordings were subsequently uploaded to the cloud-based platform, making them accessible for download and further analysis in .wav format on a personal computer (PC) or laptop.

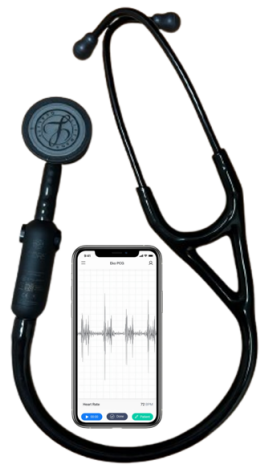

**FIGURE 4.** The digital stethoscope and the recording app.

### D. Data Processing

After transferring the recordings to a laptop, we visualized the time-domain signals which provides a clear view of the signal variations over time. We also generated time-frequency spectrograms to visualize the frequency content and signal energy over time. The resulting waveforms and spectrograms are essential for further analysis of the characteristics of the dataset and serve as a basis for future work in analyzing and interpreting the recorded data. Fig. 5 and Fig. 6 show three sample waveforms and spectrograms from the dataset, respectively.

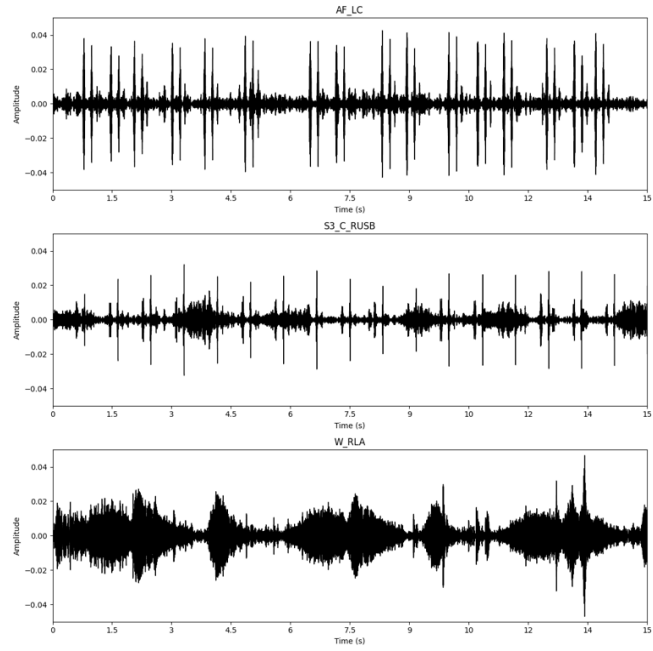

**FIGURE 5.** Time-domain waveforms of three sample recordings. From top to bottom: heart sound with atrial fibrillation recorded from the left costal margin (AF_LC), a mixture of heart and lung sounds with third heart sound (S3) and crackles recorded from the right upper sternal border (S3_C_RUSB), and lung sound with wheezes recorded from the right lower anterior (W_RLA).

## III. VALIDATION AND QUALITY

To ensure the accuracy and reliability of the collected data, we performed several validation measures across all stages of the data acquisition and processing. The nursing team assisted us in accurately identifying auscultation landmarks and ensured that clinical aspects of the data collection process were properly followed. We made recordings in a controlled, noise-free environment with the manikin in a sitting position to eliminate external noise interference. Before recording the sounds, we precisely placed the stethoscope's diaphragm at standard auscultation points to minimize artifacts. After the recording session, we performed a qualitative analysis by listening to the audio files, ensuring they aligned with the expected patterns of heart and lung sounds.

We selected a digital stethoscope with high amplification gain, active noise cancellation, and built-in frequency filters, which allowed us to capture sounds with minimal distortion. Technical specification of the stethoscope is given in Table 1. Bluetooth data transmission ensures error-free transfer to a secure database without data loss or corruption [25, 26].

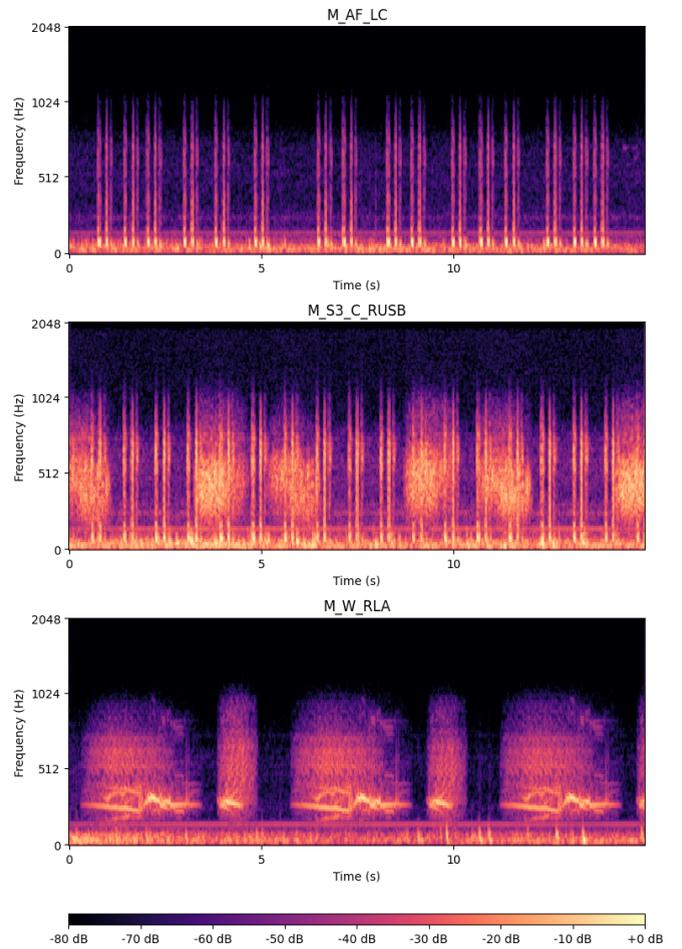

**FIGURE 6.** Time-frequency spectrogram of three sample recordings. Spectrograms are two-dimensional representations of the audio signal in Figure 5. The x-axis represents time, while the y-axis represents frequency. The signal's energy is indicated by colour. Black corresponds to areas of minimum energy, while yellow represents areas of maximum energy.

**TABLE 1.** Technical Specification of the Digital Stethoscope

| Parameter | Value |
|---|---|
| Frequency Response (Bell Mode) | [20-200] Hz [a] |
| Frequency Response (Diaphragm Mode) | [100-500] Hz [b] |
| Frequency Response (Midrange Mode) | [50-500] Hz [c] |
| Amplification | Up to 40x |
| Recording Sample Rate | 22050 Hz |
| Active Noise Cancellation | 85% [d] |

[a] Rapid decay @20 dB/octave above 300 Hz.
[b] Rapid decay @10 dB/octave below 200 Hz.
[c] Smoothed response with peak @ 550 Hz (+20 dB)
[d] @ [20 – 500] Hz, Reduction in ambient noise
dB=decibel; Hz=Hertz

Moreover, the selected manikin closely replicates real clinical scenarios and is approved by the U.S. National Council of State Boards of Nursing (NCSBN) [27, 28]. Additionally, the simulation lab team conduct regular maintenance, and perform annual calibration to ensure the manikin's sound system remains accurate over time. The reproducibility of the dataset is further supported by the standardized recording processes. Table 2 summarizes the manikin technical specifications.

**TABLE 2. Clinical Manikin Technical Specifications**

| Feature | Details |
| --- | --- |
| Dimensions | 162.56 cm x 52.07 cm x 25.4 cm |
| Approximate Weight | 22.7 kg |
| Ambient Temperature Range | 4°C to 40°C |
| Battery System | 14.4 V Lithium-ion |
| Power Adapter | 100 – 240 $V_{AC}$, 50/60 Hz |
| Simulator Network | Wireless, IEEE 802.11 g |

## IV. RECORDS AND STORAGE

The dataset consists of 210 audio recordings (101 Female, 109 Male) in .wav format, categorized into three primary groups: 50 heart sound recordings (*HS.zip*), 50 lung sound recordings (*LS.zip*), and 110 recordings of mixed heart and lung sounds (*Mix.zip*).

Each category is accompanied by a corresponding CSV file that provides metadata for the respective audio files. The CSV files (*HS.csv*, *LS.csv*, and *Mix.csv*) contain metadata about the corresponding audio files, including the file name, gender, heart and lung sound type, and the anatomical location where we recorded the sound.

The naming convention of the audio files of single recordings follows the structured format:

*Gender_Sound Type_Location.wav*

We denote the gender of the subject by *F* for female or *M* for male, followed by the sound type, and the location on the chest where we recorded the sound. For the mixture dataset, the sound type of both heart and lung is mentioned in the naming, separated by an underscore. For example, *F_LSM_R_LUSB.wav* refers to a recording of a female subject with a late systolic murmur and rhonchi, taken from the left upper sternal border.

There are ten heart sound types and six lung sound types in the dataset. Fig. 7 shows concentric donut chart of the distribution of each sound type in the dataset. We recorded sounds from twelve distinct anatomical locations. Detailed demographic information of each recording location is described in Table 3.

## V. SOURCE CODE AND SCRIPTS

The Python scripts are publicly available on GitHub at the following repository: https://github.com/Torabiy/HLS-CMDS

**TABLE 3. Chest Zones and the Number of Recordings**

| Chest Zone | | No. in *HS.zip* | No. in *LS.zip* | No. in *Mix.zip* | Total No. |
| --- | --- | --- | --- | --- | --- |
| Heart Auscultation Landmarks | RUSB | 7 | | 9 | 16 |
| | LUSB | 13 | | 11 | 24 |
| | LLSB | 10 | | 11 | 21 |
| | RC | 4 | | 13 | 17 |
| | LC | 6 | | 10 | 16 |
| | A | 10 | | 7 | 17 |
| Lung Auscultation Landmarks | RUA | | 7 | 8 | 15 |
| | LUA | | 11 | 10 | 21 |
| | RMA | | 5 | 7 | 12 |
| | LMA | | 9 | 11 | 20 |
| | RLA | | 10 | 10 | 20 |
| | LLA | | 8 | 3 | 11 |
| Total No. | | 50 | 50 | 110 | 210 |

RUSB = Right Upper Sternal Border; LUSB = Left Upper Sternal Border; LLSB = Lower Left Sternal Border; RC = Right Costal Margin; LC = Left Costal Margin; A = Apex; RUA = Right Upper Anterior; LUA = Left Upper Anterior; RMA = Right Mid Anterior; LMA = Left Mid Anterior; RLA = Right Lower Anterior; LLA = Left Lower Anterior.

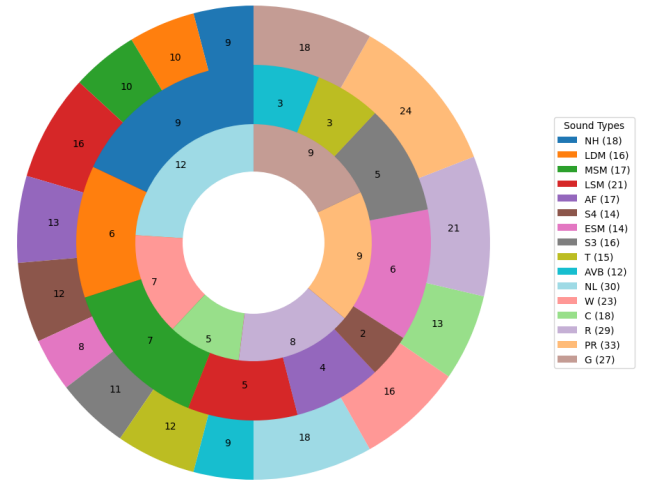

**FIGURE 7.** Concentric Donut Chart of sound types. From the outer layer to the inner level: mixed sounds (Mix.zip), heart sounds (HS.zip), and lung sounds (LS.zip). The sound type count in each folder and total count in the dataset are written. Sound Types: NH=Normal Heart; LDM=Late Diastolic Murmur; MSM=Mid Systolic Murmur; LSM=Late Systolic Murmur; AF=Atrial Fibrillation; S4=Fourth Hearth Sound; ESM=Early Systolic Murmur; S3=Third Hearth Sound; T=Tachycardia; AVB=Atrioventricular Block; NL= Normal Lung; W=Wheezing; C=Crackles; R=Rhonchi; PR=Pleural Rub; G=Gurgling.

## ACKNOWLEDGMENT

We would like to acknowledge the *Mohawk Institute for Applied Health Sciences (IAHS)* for their assistance in data collection using the patient simulators. Special thanks to *Ms. Julia Hansen* and the *Professional Practice Collaboratory (PPC)* team for providing the simulators used in our data collection process.